# Deep-Learning Identification and Global Characterization of Energetic $O_2^+$ Beams at Mars

Chi Zhang[1,*], Chuanfei Dong[1,*], Kathleen G. Hanley[2], Yaxue Dong[3], Hans Nilsson[4], Eduard Dubinin[5], Jasper Halekas[6], Janet Luhmann[2], Shannon Curry[3], Jiawei Gao[1], Yilan Qin[1], Xinmin Li[1], Hongyang Zhou[1], and Liang Wang[1]

[1]Center for Space Physics and Department of Astronomy, Boston University, Boston, MA, USA

[2]Space Sciences Laboratory, University of California, Berkeley, CA, USA

[3]Laboratory for Atmospheric and Space Physics, University of Colorado, Boulder, CO, USA

[4]Swedish Institute of Space Physics, Kiruna, Sweden

[5]Max-Planck-Institute for Solar System Research, Göttingen, Germany

[6]Department of Physics and Astronomy, University of Iowa, Iowa City, IA, USA

*Corresponding authors: Chi Zhang (zc199508@bu.edu) and Chuanfei Dong (dcfy@bu.edu)

**Key Points:**

- We developed an U-Net model for identification and statistical characterization of energetic $O_2^+$ beams at Mars.
- The beams are primarily associated with the convective electric field, while the Hall electric field can provide additional acceleration.
- The beam properties are mainly controlled by solar wind conditions, with little influence from solar EUV.

**Abstract**

Martian ionospheric ions can be accelerated away from Mars by the solar wind, contributing substantially to atmospheric ion escape. Here, we develop a U-Net-based image segmentation model to identify a distinct population of escaping planetary ions, energetic $O_2^+$ beams, from a decade of Mars Atmosphere and Volatile EvolutioN (MAVEN) observations and perform an event-based statistical analysis of their global properties and dependence on external conditions. We find that $O_2^+$ beams are widely distributed throughout the Martian plasma environment, with a strong preference for the +E hemisphere, where the solar wind convective electric field points away from Mars. This pronounced spatial asymmetry indicates that the convective electric field plays a major role in beam formation. In the solar wind and magnetosheath, these beams are consistent with the plume. Their occurrence-weighted flux agrees with previous long-term plume estimates, whereas the flux during identified beam intervals is substantially higher. This highlights the advantage of our event-based approach in capturing the intrinsic properties of $O_2^+$ plumes. In the induced magnetotail, beam formation remains strongly associated with the convective electric field, while the Hall electric field can provide additional energization. The beam properties are primarily controlled by upstream solar wind conditions, with no apparent influence from solar extreme ultraviolet irradiance. These results might improve our understanding of ion escape at Mars. More broadly, we also demonstrate the potential of deep learning for identifying transient plasma populations from spacecraft observations.

**Plain Language Summary**

Mars has lost a substantial fraction of its atmosphere to space over time. One important process is atmospheric ion escape, in which charged particles from the upper atmosphere are accelerated away from Mars by the solar wind, a stream of charged particles flowing from the Sun. Here, we use a deep-learning method called U-Net to automatically identify a distinct population of planetary oxygen ions, referred to as energetic $O_2^+$ beams, in more than a decade of observations from NASA's MAVEN mission. This allows us to directly study when and where these ion populations occur and how their properties change around Mars. We find that these $O_2^+$ beams extend from the region upstream of Mars into the disturbed plasma environment around the planet and farther into the nightside. Their density, speed, and flow direction change systematically across these regions. We also find that their properties are controlled mainly by solar wind conditions, while variations in solar ultraviolet radiation have little influence. These results provide a more complete picture of how Mars loses atmospheric ions to space and show how deep learning can help identify and study short-lived plasma features in large spacecraft data sets.

## 1 Introduction

Multiple lines of evidence indicate that Mars has lost a substantial fraction of its atmosphere to space over geological time (e.g., Jakosky et al., 2018; Lillis et al., 2015). A major contributor to this depletion is atmospheric ion escape, in which energy transferred from the solar wind accelerates planetary ions to velocities sufficient to overcome Martian gravity and escape into interplanetary space (e.g., Barabash et al., 2007; Brain et al., 2015; C. Dong et al., 2015; Luhmann & Schwingenschuh, 1990; Luhmann & Kozyra, 1991; Ramstad et al., 2018; Li et al., 2023; Zhang et al., 2021, 2025a, 2026a).

Traditionally, ion escape has been characterized through statistical analyses of long-term spacecraft observations, in which average or median ion fluxes are calculated within predefined spatial regions to reveal the global patterns of ion escape (e.g., Y. Dong et al., 2015, 2017; Dubinin et al., 2017; Inui et al., 2019; Nilsson et al., 2023). These approaches have provided a comprehensive picture of the global distribution of escaping ions and their contributions to atmospheric loss. However, such spatially averaged statistics provide limited insight into the specific processes governing ion acceleration and transport. In spacecraft observations, multiple ion populations with distinct energy and velocity distributions can coexist within the same spatial region or even at the same time, reflecting different acceleration and transport mechanisms. Averaging over these populations may therefore obscure their individual characteristics and the underlying physical processes. To better understand the formation and evolution of different ion populations, it is necessary to identify them based on their distinct observational signatures and investigate their spatial distributions and properties separately.

Among the various populations of escaping planetary ions, energetic heavy-ion beams represent a distinct population, typically characterized by relatively narrow energy distributions and energies above 100 eV. Such beam-like populations have been observed in different regions. In the upstream solar wind and magnetosheath, particularly in the +E hemisphere, where the solar wind convective electric field ($\vec{E}_{sw}$) points away from Mars, these beams are generally interpreted as the ion plume accelerated by this electric field (e.g., Carlsson et al., 2008; Ma et al., 2023;

Dubinin et al., 2006, 2023; Y. Dong et al., 2025; Liemohn et al., 2014; Qiao et al., 2026). Similar beam-like populations have also been reported in the induced magnetotail and deep wake (e.g., Dubinin et al., 2006; Fedorov et al., 2008; Nilsson et al., 2006; Zhang et al., 2024), where additional acceleration and transport processes may contribute to their formation. These observations suggest that energetic beams can arise in different plasma environments and may involve different acceleration mechanisms. However, their global occurrence, characteristic plasma properties, and dependence on external driving conditions remain poorly understood. A systematic statistical investigation is therefore needed to better characterize these beams and understand the physical processes governing their formation, acceleration, and transport at Mars.

The Mars Atmosphere and Volatile EvolutioN (MAVEN) mission (Jakosky et al., 2015) has accumulated more than a decade of observations of the Martian plasma environment, providing high-time-resolution measurements of planetary ions together with simultaneous magnetic field measurements. These comprehensive observations offer a unique opportunity to systematically investigate the global distribution, plasma properties, and potential acceleration mechanisms of energetic beams. However, identifying these beams from such an extensive data set remains challenging. Manual inspection is time-consuming and may introduce subjective biases, while simple threshold-based methods often struggle to distinguish beams from other ion populations, particularly when their spectral signatures are weak or overlap with background populations. Recent advances in deep learning, particularly image-segmentation techniques, offer a promising approach for automatically identifying distinct ion populations from complex plasma measurements (e.g., Ronneberger et al., 2015; Camporeale, 2019).

Here, we develop a U-Net-based image-segmentation method to systematically identify energetic $O_2^+$ beam events in MAVEN observations from 2015 to 2025. By isolating spectrally distinct $O_2^+$ beams from low-energy ion populations and background noise, we construct an event database that enables a long-term statistical investigation of this specific ion population. Using this database, we examine the global occurrence and spatial distribution of energetic $O_2^+$ beams, characterize their density, bulk velocity, and flux, and investigate their relationships with upstream solar wind conditions and solar EUV irradiance. We further discuss the implications of these results for the acceleration and transport of planetary ions in different plasma regions. By focusing on a distinct ion population rather than on spatially averaged ion fluxes, this study complements conventional analyses and provides new insights into planetary ion acceleration, transport, and escape at Mars.

## 2 Datasets and Coordinates

We primarily use ion composition measurements from the Suprathermal and Thermal Ion Composition (STATIC) instrument (McFadden et al., 2015) onboard the Mars Atmosphere and Volatile EvolutioN (MAVEN) mission (Jakosky et al., 2015). Specifically, we use the STATIC C6 and D1 data products. The C6 product provides omnidirectional energy spectra for different mass-to-charge channels, including $H^+$, $O^+$, and $O_2^+$, while the D1 product provides the corresponding three-dimensional ion velocity distributions. STATIC employs a time-of-flight system to measure ion mass per charge, enabling the identification and separation of different ion species. The field of view of STATIC is $360° \times 90°$.

In this study, we use background-subtracted STATIC data, in which several major instrumental background sources, including coincident events and ion straggling, have been corrected (see Hanley, 2023, for details). We also focus specifically on $O_2^+$ beams. $O_2^+$ is a major constituent of the Martian ionosphere and therefore serves as a representative tracer of planetary ions originating from the ionosphere. In addition, $O_2^+$ beams generally exhibit relatively distinct signatures in the STATIC energy spectra, making them well suited for systematic identification. However, $O^+$ can originate from both the ionosphere and the extended oxygen corona, resulting in a mixture of populations with different source regions and potentially more complex spectral signatures. We therefore restrict the present analysis to $O_2^+$ beams, while a systematic investigation of the associated $O^+$ populations is left for future work.

To investigate the dependence of the $O_2^+$ beams on external solar wind conditions, we additionally use magnetic field measurements from the MAVEN Magnetometer (MAG; Connerney et al., 2015), and solar wind measurements from the Solar Wind Ion Analyzer (SWIA; Halekas et al., 2015). To examine the influence of solar extreme ultraviolet (EUV) irradiance, we use the integrated solar irradiance in the 17–22 nm wavelength range measured by the Extreme Ultraviolet Monitor (EUVM; Eparvier et al., 2015).

We employ two Cartesian coordinate systems. The first is the Mars Solar Orbital (MSO) coordinate system, in which the $\vec{X}_{MSO}$ is along the vector from Mars to the Sun, $\vec{Z}_{MSO}$ is perpendicular to the orbital plane, and $\vec{Y}_{MSO}$ completes the right-handed system, closely aligned with the opposite direction of the orbital velocity vector. The second is the Mars Solar Electric (MSE) coordinate system. In this system, $\vec{X}_{MSE}$ points antiparallel to the upstream solar wind flow, $\vec{Z}_{MSE}$ points along the direction of the $\vec{E}_{SW}$, and the $\vec{Y}_{MSE}$ axis completes the right-handed system.

## 3 Method

We use a deep-learning-based image-segmentation approach to automatically identify energetic $O_2^+$ beams in STATIC energy spectrograms and distinguish them from other spectral populations. In this framework, each time-energy spectrogram is treated as an image, and the model determines which pixels belong to an $O_2^+$ beam based on both their local spectral characteristics and the surrounding spectral context. This allows beam structures to be identified systematically and consistently over the long-term MAVEN dataset.

For this purpose, we adopt the U-Net architecture, a convolutional neural network originally developed for image segmentation (Ronneberger et al., 2015) and now widely applied to segmentation problems in medical imaging, remote sensing, and Earth science (Siddique et al., 2021; Lv et al., 2023). Unlike a conventional image classifier, which assigns a single label to an entire image, U-Net assigns a label to each individual pixel. This is particularly well suited for plasma spectrograms, where an $O_2^+$ beam typically occupies only part of the time-energy plane and may coexist with other ion populations. By combining fine-scale spectral features with broader contextual information, the network can delineate the region occupied by each beam in the time-energy plane, thereby determining its location, duration, and energy range.

To apply the U-Net to STATIC observations, we construct the input spectrograms from the STATIC C6 measurements as follows. The STATIC C6 spectra have a temporal cadence of 4 s. Each model input consists of 512 consecutive time samples and 32 energy bins, corresponding to a time interval of 2048 s. The window length of 512 samples was selected for two reasons. First, the U-Net contains three downsampling levels, requiring the input dimensions to be divisible by $2^3 = 8$ to maintain dimensional consistency through the downsampling and upsampling operations. Second, $O_2^+$ beam intervals typically persist for approximately 10 min, so a 2048 s window provides sufficient temporal context to capture the full beam structure together with the surrounding plasma environment. We also tested shorter and longer window lengths of 256 and 1024 time samples, respectively, and obtained comparable results, indicating that the model performance is relatively insensitive to the choice of window length. Consecutive input windows overlap by 90%, corresponding to a shift of approximately 51 time samples, as illustrated at the top of Figure 1.

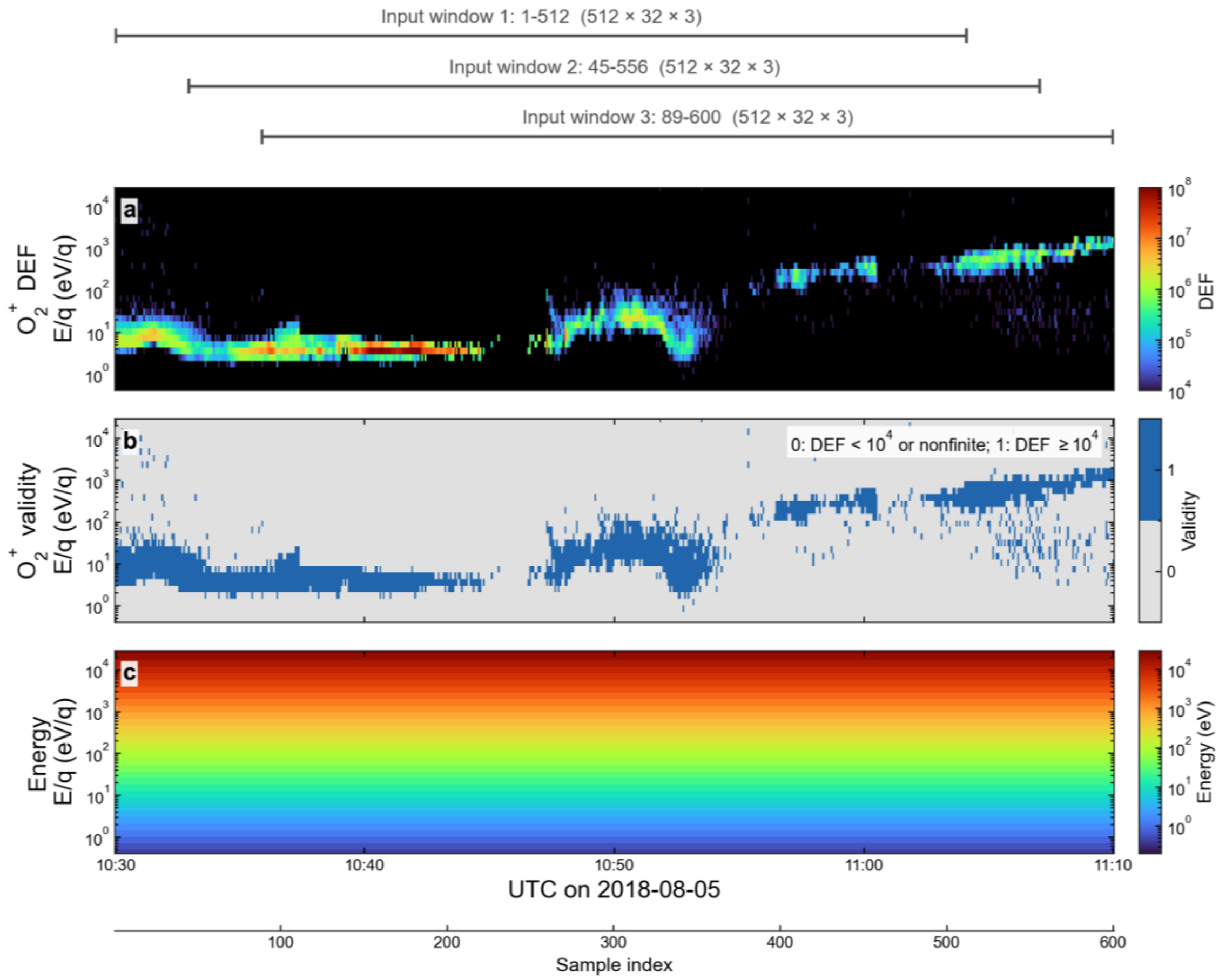

**Figure 1.** Example of the three input quantities used by the U-Net model. (a) Differential energy fluxes (DEF) of $O_2^+$. (b) Corresponding validity masks (see text for details). (c) Normalized logarithmic energy coordinate. The brackets at the top illustrate the sliding input windows, each containing 512 time samples × 32 energy bins × 3 input channels. Consecutive windows overlap by 90%, corresponding to a shift of approximately 51 samples.

To provide the network with the information needed to identify the beams, each time-energy pixel is described by three input quantities (Figure 1). The first is the differential energy fluxes (DEF) of $O_2^+$ (Figure 1a), derived using the corresponding mass ranges of 24–40 (Fowler et al., 2022). To distinguish physically meaningful signals from measurements near the instrumental noise level, a corresponding validity-mask channel is included (Figure 1b), in which measurements with $\mathrm{DEF} < 10^4\ \mathrm{keV/cm^2\ s \cdot sr \cdot keV}$ are flagged as invalid. This threshold was chosen empirically. We also tested a higher threshold of $5 \times 10^4$ and found that the results remained largely unchanged. Finally, a normalized logarithmic energy-coordinate channel (Figure 1c) provides the network with explicit information about the energy location of each spectral feature.

The U-Net architecture used in this study consists of three encoder stages, a bottleneck, and three corresponding decoder stages (Figure 2). Each convolutional block contains two successive 3×3 convolutional layers, with each convolution followed by group normalization and a rectified linear unit (ReLU) activation, as illustrated in the lower-right panel of Figure 2. In the encoder, 2×2 max-pooling progressively reduces the temporal and energy dimensions while increasing the number of feature channels. This allows the network to extract increasingly large-scale features. The bottleneck operates on the most compressed representation and captures high-level information describing the overall spectral context.

The decoder then progressively restores the original temporal and energy resolutions using transposed convolutions. At each decoder stage, the upsampled features are concatenated with the corresponding higher-resolution features from the encoder through skip connections. These connections help preserve fine-scale structures while incorporating the broader contextual information learned at deeper layers. Finally, a 1×1 convolution maps the reconstructed features to three pixel-level class probabilities: noise, low-energy ions, and beam. Each time–energy pixel is assigned to the class with the highest predicted probability. The low-energy ion class is intentionally defined broadly and may include multiple distinct ion populations. Because the present study focuses specifically on identifying $O_2^+$ beams, further separation of these low-energy populations is unnecessary and is not expected to affect the subsequent analysis.

To construct the labeled dataset, we randomly selected 200 MAVEN orbits and manually labeled two spectral populations: $O_2^+$ beams and low-energy ions. All remaining unlabeled pixels were assigned to the noise class. The labeled dataset was then divided into training and test sets, with 80% used for training and the remaining 20% reserved for independent testing.

The model was trained using the AdamW optimizer with an initial learning rate of 0.001. A cross-entropy loss function was used to quantify the difference between the predicted pixel-level class probabilities and the manually labeled classes. The learning rate was automatically reduced when the model performance stopped improving. We use the F1 score as the evaluation metric, which combines precision and recall into a single value ranging from 0 to 1, with higher values indicating better classification performance. The F1 scores for beams, noise, and low-energy ions were 0.9191, 0.9983, and 0.9797, respectively, yielding a macro F1 score of 0.9657. The slightly lower F1 score for the beam class mainly arises from weak beam signals whose fluxes are only marginally above the ambient level and are therefore classified as noise. Such misclassifications primarily remove marginal beam pixels and thus are unlikely to substantially affect the statistical results.

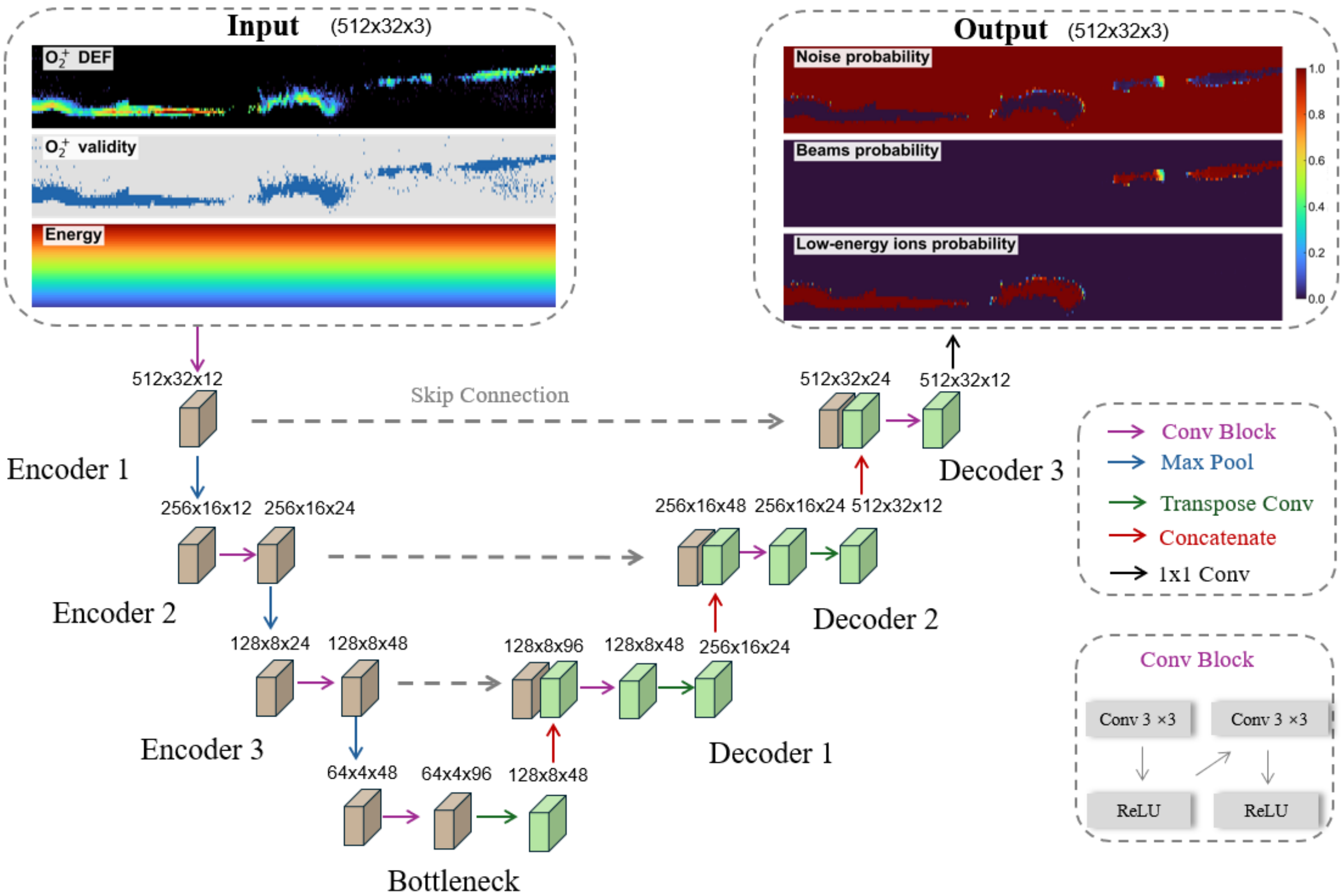

**Figure 2.** Architecture of the U-Net model. The input consists of three channels, with each input window containing 512 time samples and 32 energy bins. The encoder extracts multiscale spectral features through three convolutional and max-pooling stages, while the decoder progressively restores the original temporal and energy resolutions using transposed convolutions and skip connections. A final $1 \times 1$ convolution produces three class scores for each pixel, which are converted into probabilities for the noise, beam, and low-energy ion classes using a softmax function.

After applying the model to the full STATIC dataset, we found that some apparent $O_2^+$ signals were caused by $H^+$ contamination, in which intense proton counts produced spurious counts in the $O_2^+$ mass channel (Hanley, 2023). An example is shown in Figure S1. These contaminated signals occur at energies similar to those of $H^+$ and exhibit beam-like features, causing them to be misclassified by the model as $O_2^+$ beams (Figure S1c). To minimize this contamination, we applied an additional filtering criterion and excluded points for which both the DEF ratio $\mathrm{H^+/O_2^+} \geq 5$ and the $H^+$ DEF $\geq 10^5\ \mathrm{keV/cm^2\ s \cdot sr \cdot keV}$. After applying this filter, the false beam detections are removed (Figure S1d). This filtering criterion was therefore applied to all model-identified beam detections.

Figure 3 presents the performance of the U-Net model for observations on January 20, 2018. As indicated by the $H^+$ energy spectra in Figure 3a, MAVEN repeatedly crossed the solar wind, magnetosheath, and induced magnetosphere, with the corresponding plasma regions labeled at the top of the panel. Figure 3b shows the $O_2^+$ energy spectra, in which several distinct ion populations can be identified by visual inspection. Within the induced magnetosphere, low-energy $O_2^+$ with energies generally below 100 eV are frequently observed. In contrast, $O_2^+$ in the solar wind and magnetosheath are generally weak or absent, with low-energy populations rarely observed. Nevertheless, distinct energetic $O_2^+$ beams with energies typically above approximately 100 eV are intermittently detected in these regions. Figure 3c shows the corresponding U-Net model results. We can see that the model successfully distinguishes the low-energy ion population and $O_2^+$ beams, with the classified patterns showing good agreement with the spectral features identified by visual inspection.

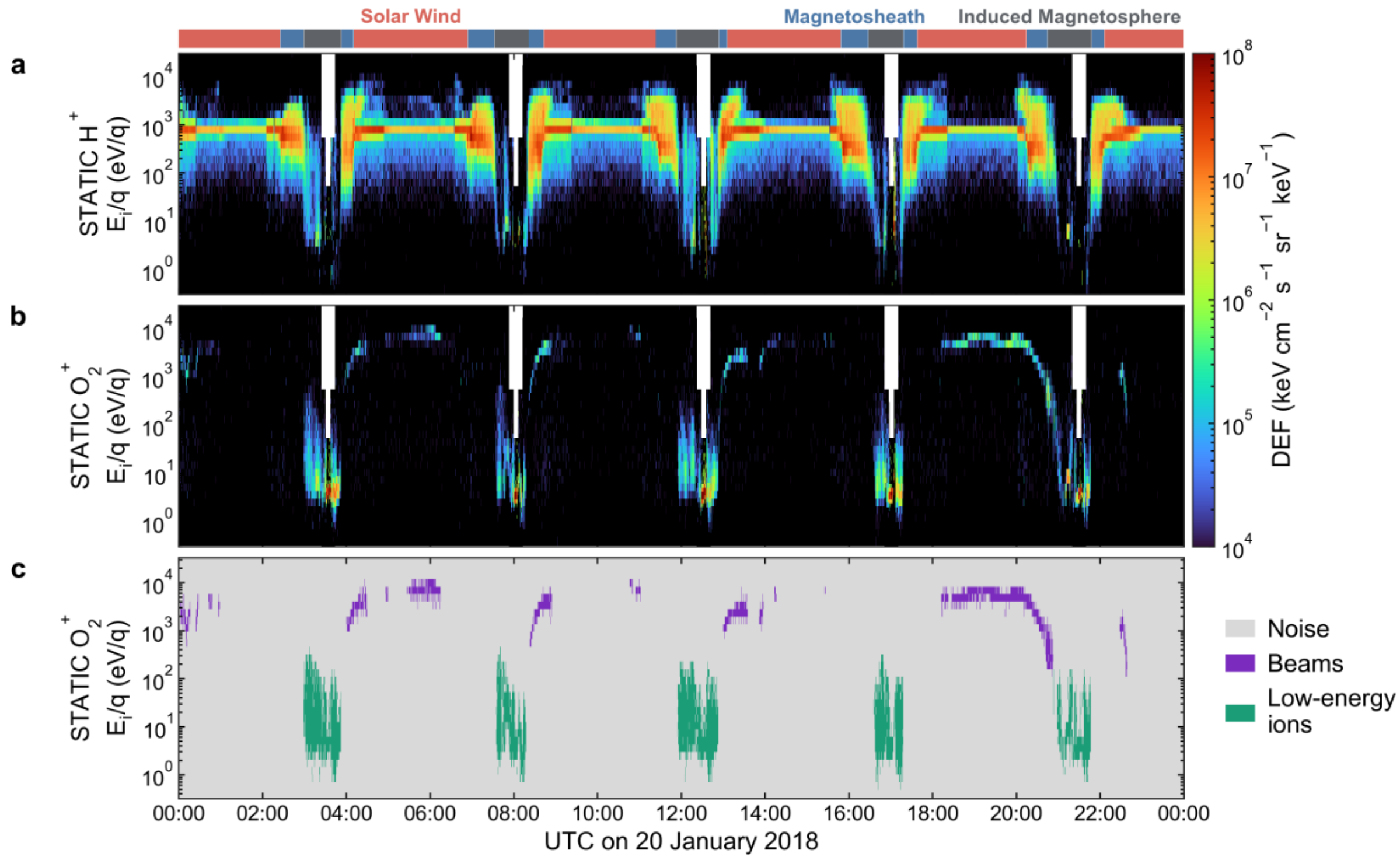

**Figure 3.** Example of energetic $O_2^+$ beam identification on 20 January 2018. (a–b) energy spectrograms of $H^+$ and $O_2^+$, respectively. The colored bars at the top indicate the corresponding plasma regions. (c) Segmented $O_2^+$ spectral populations, including noise, beams, and low-energy ions.

## 4 Results

To characterize the global properties of these energetic $O_2^+$ beams, we perform a statistical analysis of all beams identified by the U-Net. To examine the spatial distribution and properties of these beams, we further organize the observations in the MSE coordinate system. For beams observed in the upstream solar wind, the MSE coordinates are constructed directly from the simultaneous solar wind velocity measured by SWIA and the magnetic field measured by MAG. For beams observed in the magnetosheath or induced magnetosphere, the corresponding upstream solar wind conditions are determined from the inbound and outbound solar wind portions of the same MAVEN orbit, following Zhang et al. (2025b, 2026b). We retain only orbits with relatively steady upstream conditions and an angle of less than 30° between the inbound and outbound IMF vectors. The MSE coordinate system is then constructed using the average solar wind velocity and IMF. Orbits lacking suitable upstream observations on both sides are excluded because a reliable MSE coordinate system cannot be established.

After applying the model to the full dataset, the beam catalog contains 9878441 time samples. Adjacent beam intervals separated by gaps of no more than 10 min are merged, yielding 36888 intervals in total. After applying the MSE coordinate validity criteria described above, 14484 intervals containing 2811946 valid samples are retained for subsequent statistical analyses. The full list of retained intervals is publicly available in Zhang (2026c).

### 4.1 Representative Energetic $O_2^+$ Beam Events

Figure 4 shows representative energetic $O_2^+$ beam events observed in the upstream solar wind, magnetosheath, and nightside induced magnetosphere, illustrating how their characteristics vary across different plasma regions.

Figures 4a1–4e1 show an example of an $O_2^+$ beam observed in the upstream solar wind. The beam appears as a distinct high-energy population at approximately 2-10 keV and disappears at around 20:41 UTC (Figure 4b1), coincident with a clear rotation of the IMF (Figure 4d1). This rotation changes the direction of $\vec{E}_{sw}$ and shifts the +E hemisphere relative to MAVEN, moving the spacecraft from the +E hemisphere toward the −E side (Figure 4e1). The simultaneous disappearance of the beam and the transition out of the +E hemisphere are therefore consistent with the expected behavior of the classical $O_2^+$ plume: as $\vec{E}_{sw}$ rotates, the plume is displaced away from the spacecraft and is no longer detected.

Figures 4a2–4e2 show an $O_2^+$ beam observed as MAVEN moved outward through the magnetosheath toward the upstream solar wind. The beam energy progressively increases during the crossing (Figure 4b2). Together with its narrow beam-like spectral signature and its occurrence in the +E hemisphere (Figure 4e2), this behavior is consistent with the plume populations previously reported in the magnetosheath (Halekas et al., 2019; Dubinin et al., 2023). The increasing $O_2^+$ beam energy may reflect continued acceleration as the ions propagate outward from Mars. It may also be enhanced by the increasing background proton flow speed, which strengthens the local convective electric field and can provide stronger acceleration.

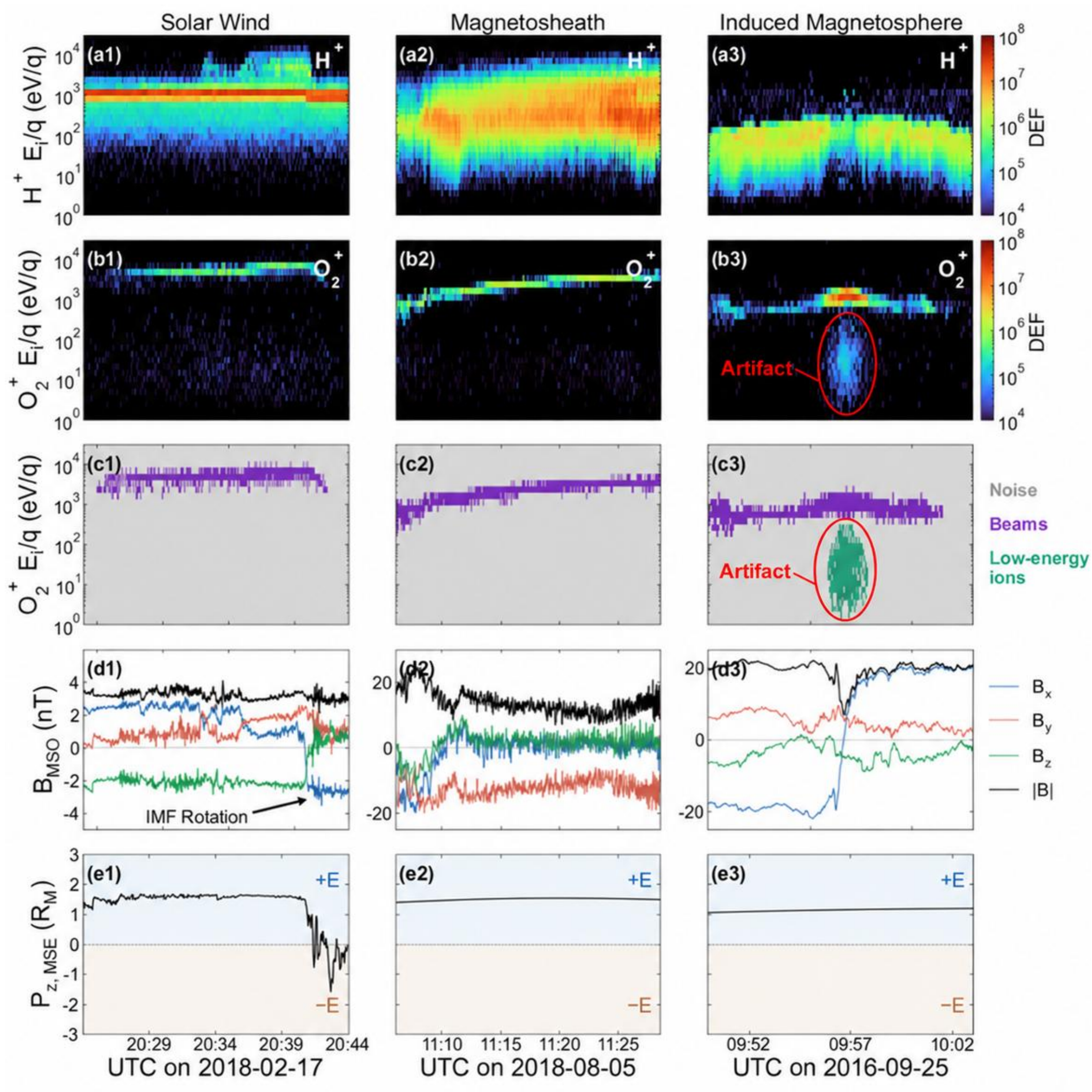

**Figure 4.** Representative energetic $O_2^+$ beams observed in the solar wind, magnetosheath, and induced magnetosphere. The three columns show examples from the solar wind (left), magnetosheath (middle), and induced magnetosphere (right), respectively. Panels (a1–a3) show the $H^+$ energy spectra, panels (b1–b3) show the $O_2^+$ energy spectra, and panels (c1–c3) show the corresponding model results. Panels (d1–d3) show the magnetic fields. Panels (e1–e3) show the spacecraft position along the $\vec{Z}_{MSE}$.

Figures 4a3–4e3 show an $O_2^+$ beam observed in the nightside induced magnetosphere. Similar to the solar wind and magnetosheath examples, the beam is also detected in the +E hemisphere (Figure 4e3), indicating that its formation still remains connected to the convective electric field. However, the nightside event also exhibits important differences from the classical plume signatures observed upstream and in the magnetosheath. In particular, the event is associated with a current sheet crossing characterized by a reversal of Bx (Figure 4d3). Notably, the $O_2^+$ beam energy increases toward the center of the current sheet and reaches its maximum there (Figure 4b3), suggesting additional energization by the Hall electric fields (e.g., Zhang et al., 2022a; Romanelli et al., 2025). These observations indicate that, although convective electric field acceleration likely contributes to the formation of $O_2^+$ beams, it is not the sole energization mechanism. It should be noted that a distinct $O_2^+$ population, primarily between 1 and 100 eV, is observed near the center of the current sheet, and is classified as low-energy ions by our model (Figure 4c3). However, further examination suggests that this low-energy signal is likely an instrumental artifact caused by enhanced sputtering and subsequent ionization of material from the spacecraft surface due to the increased flux of energetic $O_2^+$ (Hanley, 2023).

### 4.2 Spatial Occurrence Distribution of $O_2^+$ Beams

Figure 5 shows the occurrence rate of beams in different spatial slices. To calculate the occurrence rates, we first identified all valid C6 measurements. The spatial distribution of the total number of measurements is shown in Figure S2. To ensure sufficient statistical coverage, only spatial bins containing more than 100 measurements were included in the occurrence rate calculations. In the $XZ_{\mathrm{MSE}}$ plane (Figures 5a–5c), it shows a clear preference for the +E hemisphere and extends continuously from the upstream solar wind, through the magnetosheath, and into the nightside induced magnetosphere. This +E-hemisphere preference persists from the dayside to the nightside, suggesting that the convective electric field plays an important role in beam formation. The occurrence rate is highest at large positive $Z_{MSE}$, and decreases progressively from the solar wind toward the magnetosheath and induced magnetosphere.

A clear variation is also evident in the $Y_{\mathrm{MSE}}$ direction: as shown in Figures 5d-5f, the occurrence rate peaks near $Y_{\mathrm{MSE}} = 0$ and decreases toward both $\pm Y_{\mathrm{MSE}}$ flanks. We note that some $O_2^+$ beams also appear in the -E hemisphere, primarily in the magnetosheath and induced magnetotail. These detections may partly reflect uncertainties in the MSE coordinates, because the MSE frame in these regions is constructed using the orbit-averaged IMF rather than simultaneous IMF measurements. If the IMF varies substantially during an orbit, the actual direction of the convective electric field at the time of a beam observation may differ from that inferred from the orbit-averaged IMF. In contrast, in the upstream solar wind, where simultaneous solar wind and IMF measurements are available, the $O_2^+$ beams are exclusively confined to the +E hemisphere.

We define the high-occurrence region as locations where the $O_2^+$ beam occurrence rate exceeds 20%, outlined by the magenta contour in Figure 5. This region forms a largely continuous structure extending from the dayside upstream solar wind, through the magnetosheath, and into the nightside induced magnetosphere. In the nominal solar wind and magnetosheath regions, its morphology is broadly consistent with that expected for the classical plume described by Y. Dong et al. (2015) and Cao et al. (2025): $O_2^+$ are preferentially accelerated into the +E hemisphere by the convective electric field and are subsequently deflected tailward by the Lorentz force.

Within the induced magnetosphere, the high-occurrence region extends progressively deeper into the wake with increasing downtail distance and remains largely confined to $|Y_{\mathrm{MSE}}| < 1\ R_M$. This inward extension and close spatial association with the current sheet suggest that the Hall electric field may become increasingly important farther downtail and deeper into the wake. Nevertheless, the beam occurrence remains preferentially distributed in the +E hemisphere throughout the nightside region, indicating that the convective electric field continues to play an important role in organizing their large-scale transport and acceleration. The $O_2^+$ beams within the induced magnetotail therefore likely reflect the combined effects of convective and Hall electric fields.

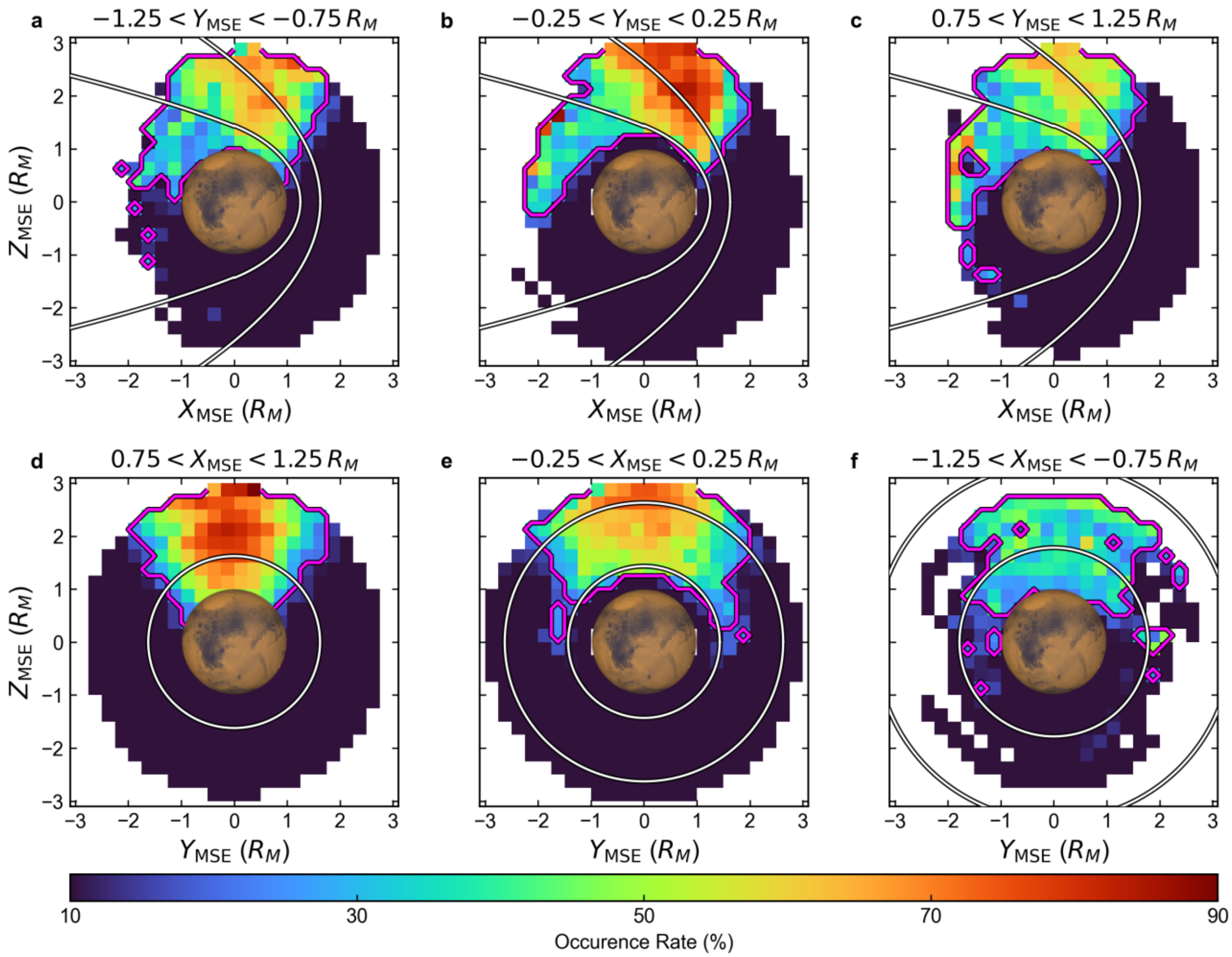

**Figure 5.** Spatial distribution of the occurrence rate of $O_2^+$ beams in MSE coordinates. (a–c) Occurrence rate in the $XZ_{\mathrm{MSE}}$ plane for three $Y_{\mathrm{MSE}}$ ranges: $-1.25 < Y_{\mathrm{MSE}} < -0.75\,R_M$, $-0.25 < Y_{\mathrm{MSE}} < 0.25\,R_M$, and $0.75 < Y_{\mathrm{MSE}} < 1.25\,R_M$, respectively. (d–f) Occurrence rate in the $YZ_{\mathrm{MSE}}$ plane for $0.75 < X_{\mathrm{MSE}} < 1.25\,R_M$, $-0.25 < X_{\mathrm{MSE}} < 0.25\,R_M$, and $-1.25 < X_{\mathrm{MSE}} < -0.75\,R_M$, respectively. The magenta contours indicate an occurrence rate of 20%. The white and black curves mark the nominal locations of the magnetic pileup boundary and bow shock, respectively, based on the empirical model of Trotignon et al. (2006).

### 4.3 Density, Bulk Velocity, and Flux

To investigate the properties of the identified $O_2^+$ beams, we calculate their plasma moments using the integration method (e.g., Fränz et al., 2006; Zhang et al., 2022b). Because the U-Net model identifies the energy range occupied by each beam, the moment calculation is restricted to the corresponding identified energy bins. Reliable moment estimates require the beam to be sufficiently observed within the STATIC field of view (FOV). We therefore examine the angular distribution of each identified beam. STATIC has an elevation-angle coverage of $-45^{\circ}$ to $+45^{\circ}$. Because these beams are generally consistent with acceleration by the convective and Hall electric fields, we retain only measurements for which both the $\vec{X}_{MSE}$ and $\vec{Z}_{MSE}$ directions lie within the FOV, corresponding to elevation angles of less than $45^{\circ}$ in absolute value in the instrument frame.

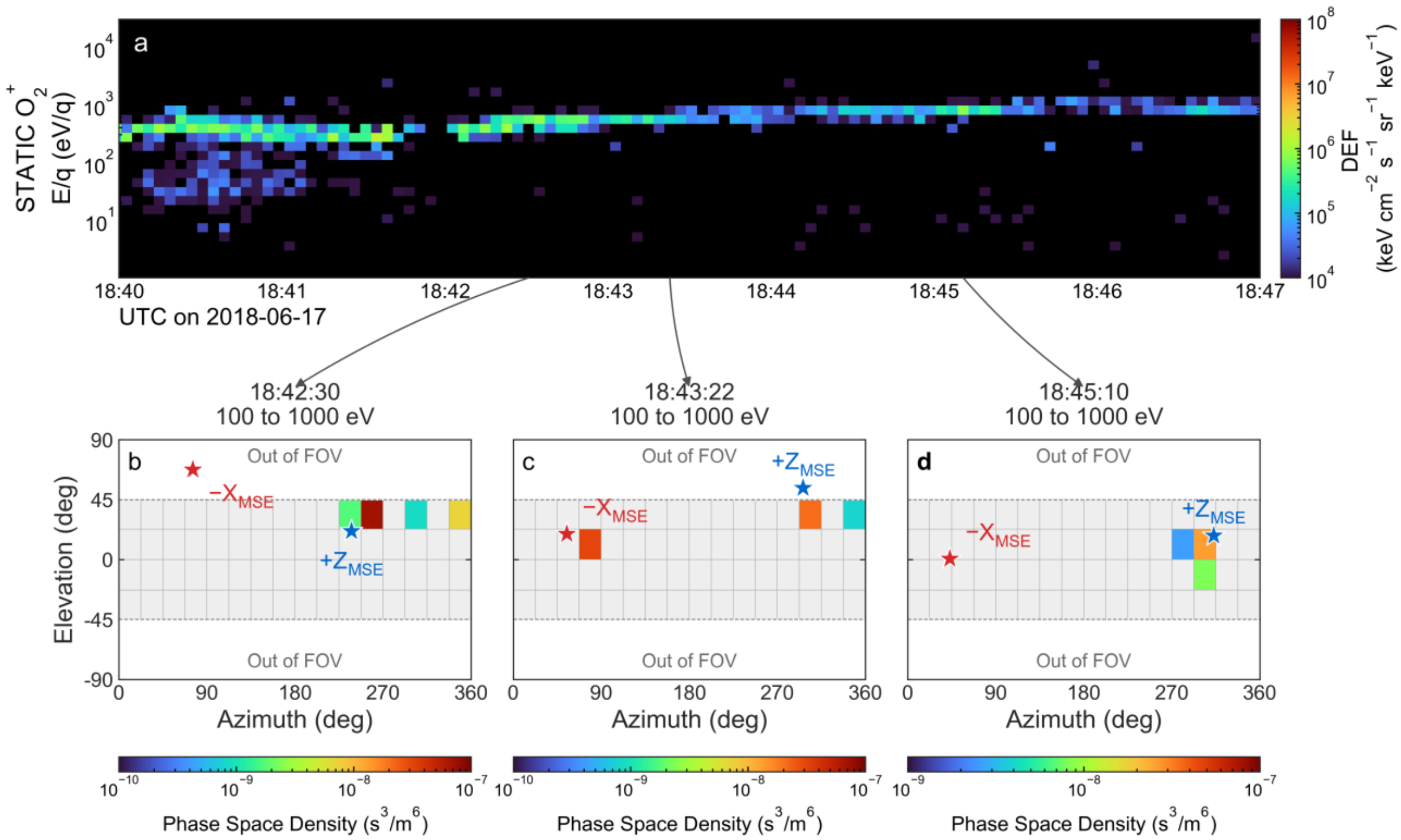

**Figure 6.** (a) $O_2^+$ energy spectrum from 18:40 to 18:47 UTC on 17 June 2018, showing an energetic $O_2^+$ beam event. (b–d) Corresponding $O_2^+$ angular distributions in STATIC instrument coordinates at 18:42:30, 18:43:22, and 18:45:10 UTC, respectively. The red and blue stars indicate the directions of $-\vec{X}_{MSE}$ and $+\vec{Z}_{MSE}$, respectively.

Figure 6 illustrates the application of this FOV criterion to an identified $O_2^+$ beam interval. Figures 6b–6d show the angular distributions at three representative times. At 18:42:30 UTC (Figure 6b), the $-\vec{X}_{MSE}$ lies outside the FOV, whereas at 18:43:22 UTC (Figure 6c), the $+\vec{Z}_{MSE}$ lies outside the FOV. In both cases, a substantial fraction of the observed $O_2^+$ population is located at elevation angles between approximately $22.5^{\circ}$ and $45^{\circ}$, close to the edge of the FOV. This suggests that the $O_2^+$ beam may extend beyond the instrument coverage and that part of the beam may therefore be missed. In contrast, at 18:45:10 UTC (Figure 6d), both the $-\vec{X}_{MSE}$ and $+\vec{Z}_{MSE}$ lie within the FOV, and the $O_2^+$ population is concentrated at elevation angles between $-22.5^{\circ}$ and $+22.5^{\circ}$, well within the FOV. The beam is therefore considered sufficiently resolved by STATIC. Plasma moments are calculated only for beam measurements that satisfy this FOV criterion. We also tested a stricter FOV criterion requiring the absolute elevation angles of both the $-\vec{X}_{MSE}$ and $+\vec{Z}_{MSE}$ to be less than 22.5°. Although this criterion substantially reduces the spatial coverage of valid measurements, the overall spatial trends remain largely unchanged.

We next examine the spatial variations in the density, bulk velocity, and flux of the identified $O_2^+$ beams. To reduce uncertainties associated with the MSE coordinate determination, we restrict the analysis to measurements in the +E hemisphere. We identified all valid D1 measurements with valid MSE coordinates, and their spatial distribution is shown in Figure S3. To ensure sufficient statistical coverage, only bins containing more than 100 measurements were included in the analysis.

Figure 7 shows the spatial distributions of $O_2^+$ beam density, bulk velocity, and flux within the high-occurrence region. The beam density exhibits a clear spatial gradient (Figures 7a and 7b), with the highest values generally observed close to Mars, particularly within the induced magnetosphere, and progressively decreasing outward through the magnetosheath into the upstream solar wind.

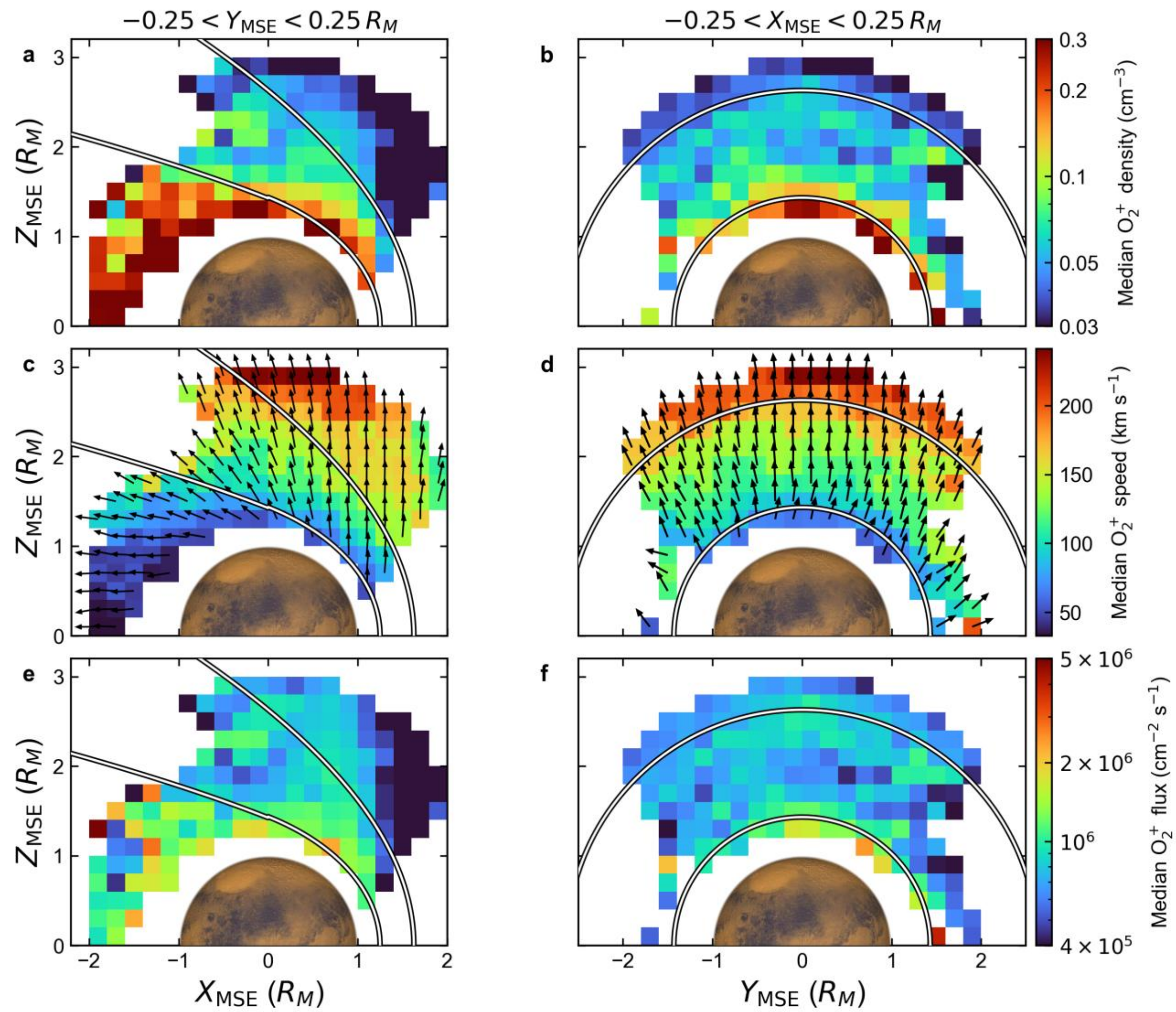

**Figure 7.** Spatial distributions of the $O_2^+$ beam plasma properties. (a, b) Median $O_2^+$ number density in the $XZ_{\mathrm{MSE}}$ and $YZ_{\mathrm{MSE}}$ planes, respectively. (c, d) Median $O_2^+$ bulk speed, with arrows indicating the corresponding velocity direction projected onto each plane. (e, f) Median $O_2^+$ flux.

The bulk velocity exhibits an opposite spatial trend (Figures 7c and 7d). In the $XZ_{\mathrm{MSE}}$ plane (Figure 7c), the $O_2^+$ beam has both tailward and +E-directed velocity components. In the upstream solar wind, the motion is dominated by the +E component, and the beam speed generally increases with increasing $Z_{\mathrm{MSE}}$, consistent with acceleration by the convective electric field. In the magnetosheath, the flow becomes increasingly tailward while the overall speed decreases. Farther downstream in the induced magnetotail, the beam motion is predominantly tailward and substantially slower. Overall, these results indicate a gradual transition from primarily +E-directed motion in the upstream solar wind to increasingly tailward transport through the magnetosheath and into the induced magnetotail.

The $YZ_{\mathrm{MSE}}$ velocity distribution (Figure 7d) further reveals the lateral transport of the $O_2^+$ beam. The velocity remains predominantly directed along $+Z_{\mathrm{MSE}}$, with the strongest $Z_{\mathrm{MSE}}$ component occurring near the $Y_{\mathrm{MSE}} = 0$ plane. Away from this central plane, however, a systematic $Y_{\mathrm{MSE}}$ component develops: $V_Y$ is generally positive in the $+Y_{\mathrm{MSE}}$ hemisphere and negative in the $-Y_{\mathrm{MSE}}$ hemisphere. This pattern indicates that the $O_2^+$ beams expand laterally away from the $Y_{\mathrm{MSE}} = 0$ plane toward both $\pm Y_{\mathrm{MSE}}$ flanks.

The flux carried by the $O_2^+$ beams (Figures 7e and 7f) reflects the combined spatial variations of density and speed. Relatively high fluxes are found closer to Mars and within the induced magnetosphere, where the enhanced density compensates for the lower bulk speed. The flux generally decreases outward toward the magnetosheath and upstream solar wind, where the $O_2^+$ beams become more tenuous despite their higher speed.

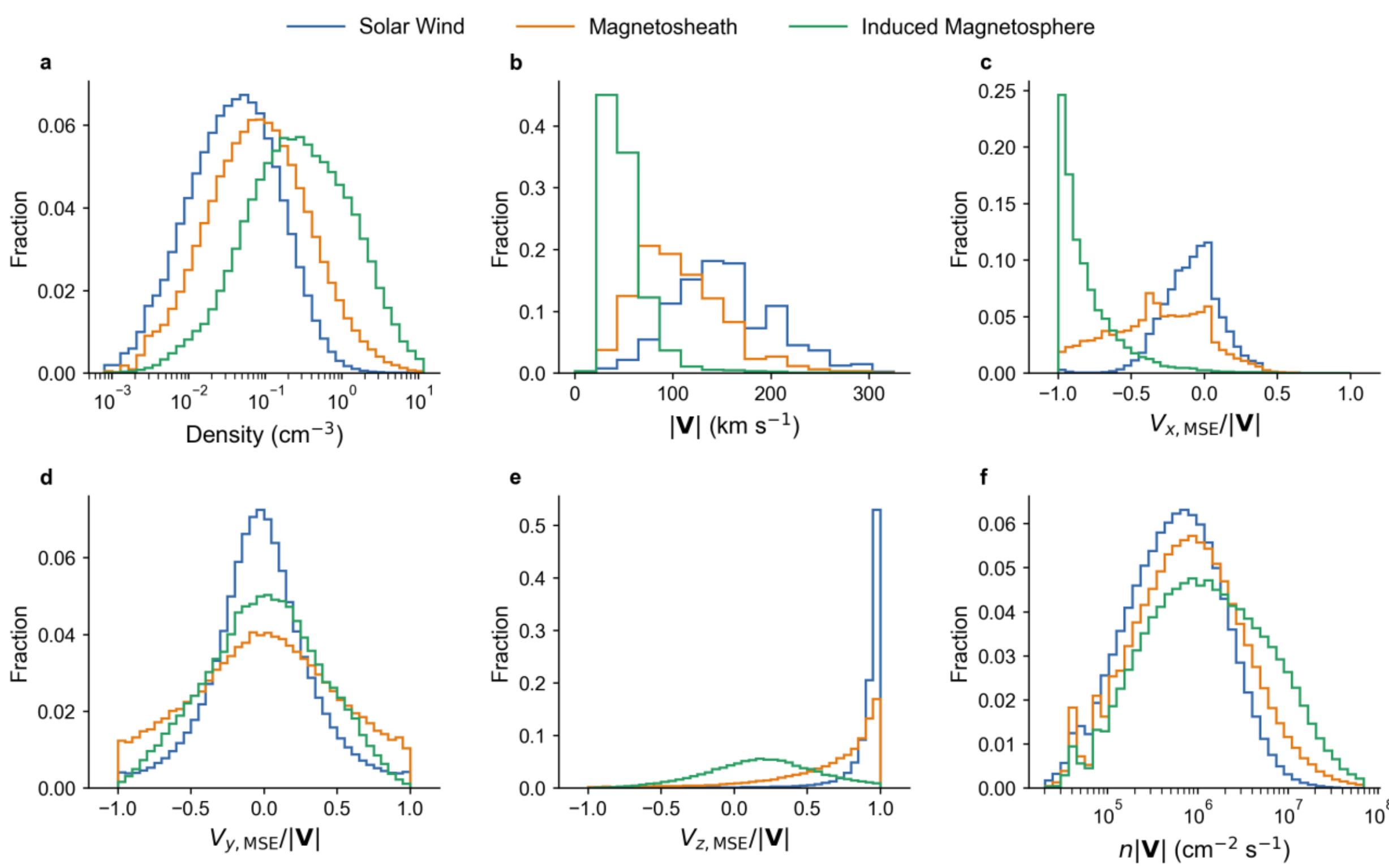

**Figure 8.** (a) $O_2^+$ number density. (b) Bulk speed, $|V|$. (c–e) Normalized bulk velocity components, $V_{X,\mathrm{MSE}}/|V|$, $V_{Y,\mathrm{MSE}}/|V|$, and $V_{Z,\mathrm{MSE}}/|V|$, respectively. (f) $O_2^+$ flux, $n|V|$. Blue, orange, and green

curves represent measurements in the solar wind, magnetosheath, and induced magnetosphere, respectively.

To more accurately compare the $O_2^+$ beam properties across different plasma regions, we use the machine-learning-based method developed by Qin et al. (2026) to identify the time intervals corresponding to the solar wind, magnetosheath, and induced magnetosphere, and then categorize the detected $O_2^+$ beams accordingly. Figure 8 provides a quantitative comparison of the $O_2^+$ beam properties among these three regions, complementing the spatial distributions shown in Figure 7. The density distributions (Figure 8a) shift systematically toward higher values from the solar wind to the magnetosheath and induced magnetosphere, with characteristic values of approximately 0.053 $cm^{-3}$, 0.086 $cm^{-3}$, and 0.284 $cm^{-3}$, respectively.

The total $O_2^+$ beam speed, $|V|$ show opposite trend, it decreases substantially from the solar wind toward the induced magnetosphere (Figure 8b), with the distribution shifting from higher speeds in the solar wind to progressively lower speeds in the magnetosheath and induced magnetosphere. The normalized velocity components further reveal a systematic change in the beam flow direction (Figures 8c-8e). In the solar wind, the beam motion is strongly dominated by the $+Z_{\mathrm{MSE}}$ component, with $V_{z,\mathrm{MSE}}/|V|$ peaking near +1, whereas $V_{x,\mathrm{MSE}}/|V|$ peaks near -0.1. In the induced magnetosphere, this pattern is largely reversed: $V_{x,\mathrm{MSE}}/|V|$ peaks near -1, indicating predominantly tailward flow. The magnetosheath exhibits an intermediate distribution between these two regimes. In contrast, $V_{y,\mathrm{MSE}}/|V|$ remains broadly centered near zero in all three regions. Together, these distributions quantitatively demonstrate a transition from fast, predominantly +E-directed motion in the upstream solar wind to slower, increasingly tailward motion toward the induced magnetotail.

Despite the differences in density and speed, the flux distributions overlap substantially among the three regions (Figure 8f). Their peak values are of the same order, approximately $7.09 \times 10^5$, $8.69 \times 10^5$, and $8.69 \times 10^5$ $cm^{-2}s^{-1}$ in the solar wind, magnetosheath, and induced magnetosphere, respectively. However, the high-flux tail progressively extends toward larger values from the solar wind to the magnetosheath and induced magnetosphere. To compare our event-based $O_2^+$ beam fluxes with the long-term average $O_2^+$ plume flux reported by Y. Dong et al. (2023), we apply the same spatial selection used in that study: $1.2 \leq Z_{\mathrm{MSE}} \leq 2.0\, R_M$, $X_{\mathrm{MSE}}^2 + Y_{\mathrm{MSE}}^2 \leq (2.5\, R_M)^2$, and $X_{\mathrm{MSE}} \geq -1.6\, R_M$. Within this region, the median $O_2^+$ beam flux is $6.87 \times 10^5\ \mathrm{cm^{-2}\ s^{-1}}$, more than twice the long-term average value of $3.0 \times 10^5\ \mathrm{cm^{-2}\ s^{-1}}$ reported by Y. Dong et al. (2023). However, the beam occurrence rate in the same region is only 41%. Weighting the median beam flux by this occurrence rate yields an effective time-averaged flux of $2.82 \times 10^5$ cm$^{-2}$ s$^{-1}$, in close agreement with the average plume flux.

This agreement strongly supports the interpretation that the $O_2^+$ beams identified in this upstream solar wind and magnetosheath region correspond to the classical plume population. It also suggests that long-term averages based on predefined spatial regions include a substantial fraction of measurements obtained during periods of plume absence. As a result, such regional averaging can underestimate the characteristic flux carried by the plume when it is actually observed.

### 4.4 Solar Wind and EUV Dependence of $O_2^+$ Beam Properties

To investigate how these energetic $O_2^+$ beam properties vary with upstream solar wind and solar EUV conditions, we examine their relationships with the solar wind dynamic pressure, $P_{dyn} = n_{sw} m_p V_{sw}^2$, where $n_{sw}$ and $m_p$ are the solar wind proton number density and proton mass, respectively; solar wind convective electric field, $|E_{\rm sw}|$ ; and the solar EUV irradiance. Because these parameters are not independent of one another, we apply a partial Spearman rank correlation analysis to quantify the relationship between each upstream parameter and each beam property while controlling for the effects of the other parameters (Baba et al., 2004). The resulting partial Spearman correlation coefficients ($r_s$) are shown in Figure 9. Here, correlations with $|r_s| < 0.1$ are considered negligible for the following discussion.

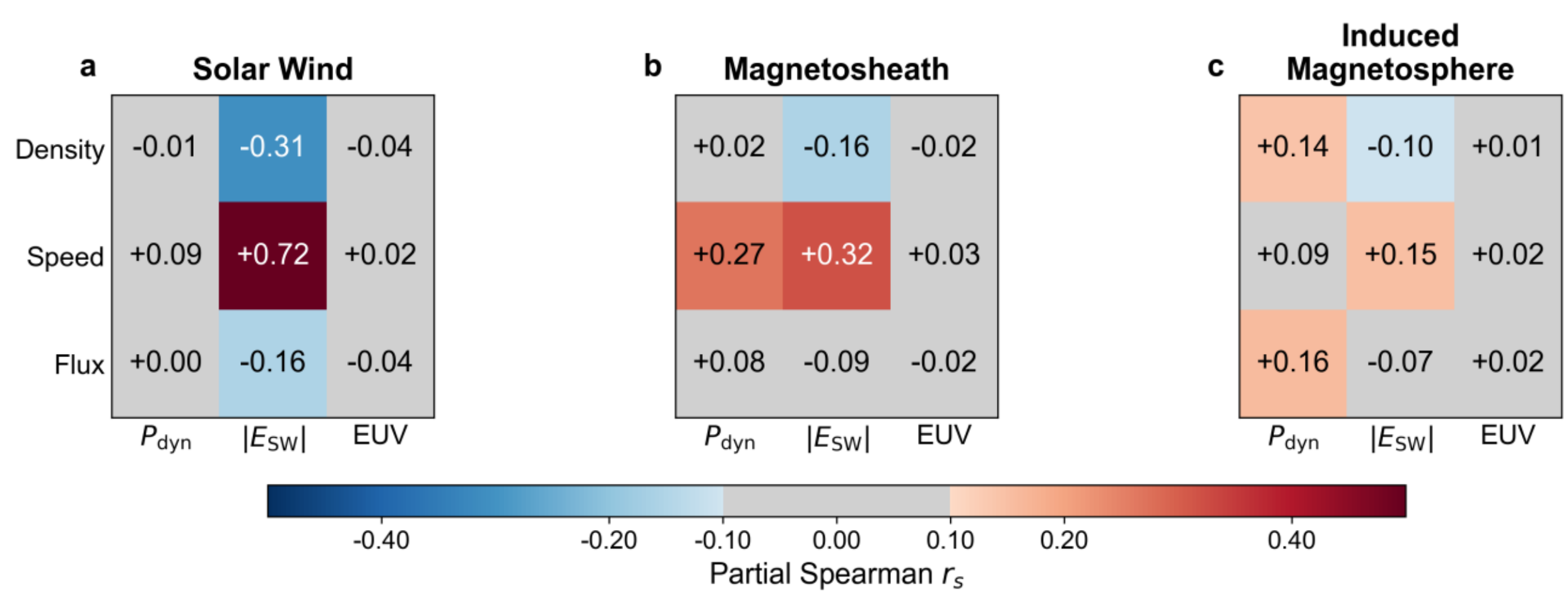

**Figure 9.** Partial Spearman correlation coefficients between $O_2^+$ beam properties and upstream solar wind parameters in the (a) solar wind, (b) magnetosheath, and (c) induced magnetosphere. The $O_2^+$ beam properties include density, speed, and flux, while the controlling parameters are the $P_{dyn}$, $|E_{sw}|$, and the solar EUV radiation.

In the upstream solar wind (Figure 9a), $|E_{\mathrm{sw}}|$ shows by far the strongest relationship with the $O_2^+$ beam properties. The beam speed is strongly positively correlated with $|E_{\mathrm{sw}}|(r_s = +0.72)$, suggesting that these upstream $O_2^+$ beams are primarily accelerated by $\vec{E}_{sw}$. In contrast, the beam density decreases with increasing $|E_{\mathrm{sw}}|$ ($r_s = -0.31$), resulting in a weaker negative correlation between flux and $|E_{\mathrm{sw}}|(r_s = -0.16)$. This inverse density dependence may partly arise from stronger acceleration: for a comparable ion flux, acceleration to higher velocities reduces the local ion density. In addition, the contribution of the IMF strength to $|E_{\mathrm{sw}}|$ may also play a role. A larger $|E_{\mathrm{sw}}|$ is generally associated with a stronger IMF, which reduces the ion gyroradius and fewer ions can reach the upstream solar wind. Thus, although a stronger $|E_{\mathrm{sw}}|$ accelerates the $O_2^+$ beams to higher speeds, the accompanying reduction in density can offset this increase, leading to the weaker negative dependence of flux on $|E_{\mathrm{sw}}|$. In comparison, $P_{\mathrm{dyn}}$ shows essentially no relationship with density or speed and only a negligible correlation with flux. Solar EUV irradiance is likewise essentially uncorrelated with all three properties.

In the magnetosheath (Figure 9b), the dependence on upstream conditions is weaker than in the solar wind but remains evident for the beam speed. Both $P_{\mathrm{dyn}}$ and $|E_{\mathrm{sw}}|$ are positively correlated with speed, with $r_s = +0.27$ and $+0.32$, respectively, suggesting that stronger upstream solar wind forcing is associated with faster $O_2^+$ beams in the magnetosheath. The dependence on the $P_{\mathrm{dyn}}$ may partly reflect the stronger magnetic fields associated with enhanced solar wind dynamic pressure in the magnetosheath, which can in turn strengthen the local convective electric field. The density remains negatively correlated with $|E_{\mathrm{sw}}|$ ($r_s = -0.16$), although the relationship is substantially weaker than in the upstream solar wind. Solar EUV again shows negligible correlations with density, speed, and flux.

In the induced magnetosphere (Figure 9c), the $O_2^+$ beam properties show only weak dependence on the upstream conditions. $P_{\mathrm{dyn}}$ is weakly positively correlated with density ($r_s = +0.14$) and flux ($r_s = +0.16$), while its relationship with speed is negligible. This suggests that the weak enhancement of flux under higher dynamic pressure is primarily associated with increased density rather than increased bulk speed. The influence of $|E_{\rm sw}|$ is also substantially reduced compared with that in the solar wind and magnetosheath: it shows only a weak positive correlation with speed ($r_s = +0.15$) and a weak negative correlation with density ($r_s = -0.10$), while its correlation with flux is negligible, possibly reflecting an increased contribution from ions accelerated by the Hall electric field. Solar EUV irradiance again shows essentially no relationship with any of the beam properties.

Overall, the influence of the upstream solar wind on the $O_2^+$ beams becomes progressively weaker from the solar wind through the magnetosheath to the induced magnetosphere. The most prominent relationship is the strong dependence of the beam speed on $|E_{\rm sw}|$ in the upstream solar wind, consistent with direct acceleration by the convective electric field. This dependence weakens substantially after the ions enters the magnetosheath and induced magnetosphere. In contrast, $P_{\rm dyn}$ becomes relatively more important in the magnetosheath and induced magnetosphere, although through different pathways. In the magnetosheath, higher $P_{\rm dyn}$ is mainly associated with increased speed, which contributes to the enhanced flux. In the induced magnetosphere, however, $P_{\rm dyn}$ shows little relationship with speed but a clearer positive correlation with density, indicating that the corresponding increase in flux is primarily density driven. Solar EUV irradiance, by comparison, shows no appreciable relationship with density, speed, or flux in any of the three plasma regions.

## 5 Conclusion and Discussion

In this study, we developed a deep-learning-based image-segmentation method to automatically identify Martian energetic $O_2^+$ beams from a decade of MAVEN observations and used the resulting event database to characterize their global properties and responses to external conditions. By directly isolating $O_2^+$ beam intervals from the observations, this event-based approach provides a more direct characterization of their intrinsic properties. The main findings are summarized as follows:

(1) The identified $O_2^+$ beams are widely distributed, extending from the upstream solar wind through the magnetosheath and into the induced magnetotail, with a strong statistical preference for the +E hemisphere. This spatial asymmetry suggests that their formation is strongly controlled by the convective electric field. In the upstream solar wind and magnetosheath, the beams largely exhibit the characteristics of the classical $O_2^+$ plume. In the nightside induced magnetotail, however, the convective electric field remains important, while the Hall electric field may provide additional energization. Farther downtail, the region of high beam occurrence progressively shifts toward the wake center, suggesting an increasingly important role of the Hall electric field in shaping and accelerating the nightside $O_2^+$ beams.

(2) The $O_2^+$ beams exhibit systematic regional variations in their plasma properties. Their density increases from the upstream solar wind through the magnetosheath to the induced magnetosphere, whereas their bulk speed decreases substantially, while the ion flux remains of the same order of magnitude across the three regions. The $O_2^+$ beam flow also shows a systematic dependence on $Y_{\mathrm{MSE}}$, with the beams preferentially directed away from the $Y_{\mathrm{MSE}} = 0$ plane toward both flanks, indicating a statistically divergent flow pattern. Within the spatial region considered by Y. Dong et al. (2023), the occurrence-weighted flux of the identified $O_2^+$ beams closely agrees with their long-term plume flux estimate, providing further evidence that these $O_2^+$ beams represent the classical plume in this region. At the same time, the median flux during the identified beam intervals is substantially higher. This difference highlights an important advantage of the event-based approach: by isolating times when the plume is actually present, it provides a more representative measure of the intrinsic plume properties, whereas previous long-term statistical estimates average over both plume-present and plume-absent intervals and therefore yield lower mean fluxes.

(3) The $O_2^+$ beam properties are controlled primarily by upstream solar wind conditions rather than by solar EUV irradiance. In the upstream solar wind, beam speed shows its strongest dependence on $|E_{\mathrm{sw}}|$, consistent with the convective electric field acceleration, while the beam density decreases with increasing $|E_{\mathrm{sw}}|$. The dependence on $|E_{\mathrm{sw}}|$ becomes progressively weaker in the magnetosheath and induced magnetosphere. In contrast, higher $P_{\mathrm{dyn}}$ is associated with enhanced flux in both the magnetosheath and induced magnetosphere, but through different pathways: in the magnetosheath, the flux enhancement is mainly associated with increased speed, whereas in the induced magnetosphere it is primarily driven by increased density. Solar EUV irradiance shows no appreciable relationship with the beam density, speed, or flux in any of the three plasma regions.

The acceleration processes responsible for the beams in the magnetotail are likely more complex than those of the classical plume in the upstream solar wind and magnetosheath. Numerical simulations have also reproduced beam-like $O_2^+$ populations in the +E hemisphere of the induced magnetotail. Hybrid simulations, which include Hall effects, produce such populations in this region (Jarvinen et al., 2018), while the test-particle simulations of Curry et al. (2014) reproduce +E-hemisphere $O_2^+$ beams using the convective electric field without including the Hall electric field. The latter result indicates that the convective electric field alone can contribute substantially to the formation of nightside $O_2^+$ beams. Our observations then further suggest that the Hall electric field may provide additional energization: the $O_2^+$ beam energy increases toward the current sheet center, where Hall effects are expected to become more important. Together, these results suggest that the convective electric field plays an important role in forming the nightside $O_2^+$ beams, while the Hall electric field may further accelerate them. Quantifying their relative contributions will require further analysis using Hall-inclusive simulations.

We propose a scenario to explain the systematic evolution of the $O_2^+$ beam properties and associated acceleration processes from the upstream solar wind toward the magnetosheath and induced magnetosphere. In a multi-ion description, solar wind protons and planetary $O_2^+$ ions can move relative to one another (Dubinin et al., 2018). In the weak mass-loading limit, where the Hall term can be neglected as a first approximation, the force accelerating $O_2^+$ in the +E direction can be approximated as, $F = eE_{sw}n_{sw}/(n_{sw} + n_{O_2})$, where $n_{sw}$ and $n_{O_2}$ are the proton and $O_2^+$ densities, respectively. As the $O_2^+$ density increases from the solar wind toward the magnetosheath, the proton fraction $n_{sw}/(n_{sw} + n_{O_2})$ decreases, weakening the effective acceleration of $O_2^+$ along the +E direction. At the same time, the increasing $O_2^+$ density strengthens mass loading, decelerates the solar wind, and enhances magnetic field draping, making the Hall electric field increasingly important. The Hall electric field can then provide additional acceleration of $O_2^+$ in the tailward direction. In addition, once accelerated, $O_2^+$ undergo gyromotion in the magnetic field, which can rotate their velocity from the +E direction toward the tailward direction. Together, these effects provide a natural explanation for the observed rotation of the $O_2^+$ beam flow from the predominantly +E-directed in the upstream region to increasingly tailward-directed within the induced magnetosphere. A similar transition occurs in cometary solar wind interactions, where plasma dynamics depend strongly on the level of outgassing and mass loading (Goetz et al., 2022). Under intermediate mass-loading conditions, newly created cometary ions exhibit pronounced finite-gyroradius, pickup-like motion controlled by the convective electric field, whereas stronger mass loading produces increasingly fluid-like behavior and predominantly tailward transport of cometary plasma. The evolution observed at Mars may represent an analogous transition.

The dependence of the $O_2^+$ beam flux on solar wind parameters, together with the lack of a clear dependence on EUV, suggests that escape associated with these $O_2^+$ beams may be primarily energy-limited rather than source-limited. In other words, the escaping efficiency of $O_2^+$ beams appears to be controlled more strongly by the available solar wind energy for ion acceleration than by the availability of ionospheric source particles. In contrast, Ramstad & Barabash (2021) suggested that the total ion escape from Mars tends toward a source-limited regime, in which further increases in the available energy do not produce a corresponding increase in the escape rate because the supply of escaping particles becomes limiting. Previous studies have also shown that magnetotail ion escape exhibits a pronounced dependence on solar EUV, whereas plume escape shows a much weaker EUV dependence (Y. Dong et al., 2017; Nilsson et al., 2023). Given that the $O_2^+$ beams observed in the magnetosheath and solar wind in our study are consistent with the classical plume population, these results indicate an important difference between the two escape channels: plume escape appears to be more strongly energy-limited, whereas magnetotail escape may be more strongly influenced by the available ionospheric source population.

Beyond its application to the ion beam, this study demonstrates the broader potential of image-segmentation-based deep learning for space-physics research. Many plasma phenomena are recognized through characteristic morphological signatures in time-energy or time-frequency spectrograms, often without a simple quantitative definition or explicit selection criteria. This makes manual event identification time-consuming and limits the feasibility of long-term statistical studies. Once a representative training set is available, image-segmentation methods such as U-Net can automate this process and enable systematic, event-based analyses across large datasets. Such an approach therefore has broad applicability for identifying and characterizing transient phenomena throughout space physics.

## Data and Code Availability Statement

The research described in this manuscript uses publicly available data from the MAVEN mission, including observations from the MAG, SWIA, and STATIC instruments (Connerney, 2026; Halekas, 2026; McFadden, 2026). The U-Net model architecture, trained model parameters, inference code, and example scripts are publicly available at https://doi.org/10.5281/zenodo.23512278 (Zhang, 2026c). The list of beam intervals identified in this study is publicly available through Zenodo at https://doi.org/10.5281/zenodo.23162891 (Zhang, 2026d).

## Acknowledgments

This work was partially supported by NASA grant NNH10CC04C through the MAVEN Project, NASA grants 80NSSC23K0911 and 80NSSC24K1843, and the Alfred P. Sloan Research Fellowship.

## Conflict of Interest Statement

The authors have no conflicts of interest to disclose.